# THE IMPACT OF QCD AND LIGHT-CONE QUANTUM MECHANICS ON NUCLEAR PHYSICS*

STANLEY J. BRODSKY

*Stanford Linear Accelerator Center*
*Stanford University, Stanford, California 94309*

and

FELIX SCHLUMPF

*Department of Physics*
*University of Maryland, College Park, Maryland 20742*

## ABSTRACT

We discuss a number of novel applications of Quantum Chromodynamics to nuclear structure and dynamics, such as the reduced amplitude formalism for exclusive nuclear amplitudes. We particularly emphasize the importance of light-cone Hamiltonian and Fock State methods as a tool for describing the wavefunctions of composite relativistic many-body systems and their interactions. We also show that the use of covariant kinematics leads to nontrivial corrections to the standard formulae for the axial, magnetic, and quadrupole moments of nucleons and nuclei.

Review talk given at the International School of Nuclear Physics
**Electromagnetic Probes and the Structure of Hadrons and Nuclei**
Erice, Sicily, September 15–23, 1994

---

★ Work supported by the Department of Energy, contract DE–AC03–76SF00515.

# INTRODUCTION

In principle, quantum chromodynamics can provide a fundamental description of hadron and nuclei structure and dynamics in terms of elementary quark and gluon degrees of freedom. In practice, the direct application of QCD to nuclear phenomena is extremely complex because of the interplay of nonperturbative effects such as color confinement and multi-quark coherence. Despite these challenging theoretical difficulties, there has been substantial progress in identifying specific QCD effects in nuclear physics. A crucial tool in these analyses is the use of relativistic light-cone quantum mechanics and Fock state methods in order to provide a tractable and consistent treatment of relativistic many-body effects. In some applications, such as exclusive nuclear processes at large momentum transfer, one can make first-principle predictions using factorization theorems which separate hard perturbative dynamics from the nonperturbative physics associated with hadron or nuclear binding. In other applications, such as the passage of hadrons through nuclear matter and the calculation of the axial, magnetic, and quadrupole moments of light nuclei, the QCD description provides new insights which go well beyond the usual assumptions of traditional nuclear physics.

In these lectures, we will outline a number of novel applications of QCD and light-cone quantum mechanics to nuclear structure and dynamics. We will particularly emphasize the importance of light-cone Hamiltonian and Fock State methods as a tool to consistently describe composite relativistic many-body systems and their electromagnetic interactions. Further discussions and references may be found in the review (Brodsky and Lepage, 1989).

# LIGHT-CONE METHODS IN QCD

In recent years quantization of quantum chromodynamics at fixed light-cone time $\tau = t - z/c$ has emerged as a promising method for solving relativistic bound-state problems in the strong coupling regime including nuclear systems (Brodsky *et al.*, 1993). Light-cone quantization has a number of unique features that make it appealing, most notably, the ground state of the free theory is also a ground state of the full theory, and the Fock expansion constructed on this vacuum state provides a complete relativistic many-particle basis for diagonalizing the full theory. The light-cone wavefunctions $\psi_n(x_i, k_{\perp i}, \lambda_i)$, which describe the hadrons and nuclei in terms of their fundamental quark and gluon degrees of freedom, are frame-independent. The essential variables are the boost-invariant light-cone momentum fractions $x_i = p_i^+/P^+$, where $P^\mu$ and $p_i^\mu$ are the hadron and quark or gluon momenta, respectively, with $P^\pm = P^0 \pm P^z$. The internal transverse momentum variables $\vec{k}_{\perp i}$ are given by $\vec{k}_{\perp i} = \vec{p}_{\perp i} - x_i \vec{P}_\perp$ with the constraints $\sum \vec{k}_{\perp i} = 0$ and $\sum x_i = 1$. *i.e.*, the light-cone momentum fractions $x_i$ and $\vec{k}_{\perp i}$ are relative coordinates, and they describe the hadronic system independent of its total four momentum $p^\mu$. The entire spectrum of hadrons and nuclei and their scattering states is given by the set of eigenstates of the light-cone Hamiltonian $H_{LC}$ of QCD. The Heisenberg problem takes the form:

$$H_{LC}|\Psi\rangle = M^2|\Psi\rangle.$$

For example, each hadron has the eigenfunction $|\Psi_H\rangle$ of $H_{LC}^{QCD}$ with eigenvalue $M^2 = M_H^2$. If we could solve the light-cone Heisenberg problem for the proton in QCD, we could then expand its eigenstate on the complete set of quark and gluon eigensolutions $|n\rangle = |uud\rangle, |uudg\rangle \cdots$ of the free Hamiltonian $H_{LC}^0$ with the same global quantum numbers:

$$|\Psi_p\rangle = \sum_n |n\rangle \psi_n(x_i, k_{\perp_i}, \lambda_i).$$

The $\psi_n$ $n = 3, 4, \ldots$ are first-quantized amplitudes analogous to the Schrödinger wavefunction, but it is Lorentz-frame independent. Particle number is generally not conserved in a relativistic quantum field theory. Thus each eigenstate is represented as a sum over Fock states of arbitrary particle number. Thus in QCD each hadron is expanded as second-quantized sums over fluctuations of color-singlet quark and gluon states of different momenta and number. The coefficients of these fluctuations are the light-cone wavefunctions $\psi_n(x_i, k_{\perp i}, \lambda_i)$. The invariant mass $\mathcal{M}$ of the partons in a given Fock state can be written in the elegant form $\mathcal{M}^2 = \sum_{i=1}^3 \frac{\vec{k}_{\perp i}^2 + m^2}{x_i}$. The dominant configurations in the wavefunction are generally those with minimum values of $\mathcal{M}^2$. Note that except for the case $m_i = 0$ and $\vec{k}_{\perp i} = \vec{0}$, the limit $x_i \to 0$ is an ultraviolet limit; *i.e.* it corresponds to particles moving with infinite momentum in the negative $z$ direction: $k_i^z \to -k_i^0 \to -\infty$.

In the case of QCD in one space and one time dimensions, the application of discretized light-cone quantization (DLCQ) (Brodsky and Pauli, 1991) provides complete solutions of the theory, including the entire spectrum of mesons, baryons, and nuclei, and their wavefunctions (Hornbostel, Brodsky, and Pauli, 1990). In the DLCQ method, one simply diagonalizes the light-cone Hamiltonian for QCD on a discretized Fock state basis. The DLCQ solutions can be obtained for arbitrary parameters including the number of flavors and colors and quark masses. More recently, DLCQ has been applied to new variants of QCD(1+1) with quarks in the adjoint representation, thus obtaining color-singlet eigenstates analogous to gluonium states (Demeterfi, Klebanov, and Bhanot, 1994).

The DLCQ method becomes much more numerically intense when applied to physical theories in 3 + 1 dimensions; however, progress is being made. An analysis of the spectrum and light-cone wavefunctions of positronium in QED(3+1) is given in (Krautgartner, Pauli, and Wolz, 1992). Currently, Hiller, Okamoto, and Brodsky (Hiller *et al.*, 1994) are pursuing a nonperturbative calculation of the lepton anomalous moment in QED using this method. Burkardt has recently solved scalar theories with transverse dimensions by combining a Monte Carlo lattice method with DLCQ (Burkardt, 1994).

Given the light-cone wavefunctions $\{\psi_n(x_i, k_{\perp_i}, \lambda_i)\}$, one can compute the electromagnetic and weak form factors from a simple overlap of light-cone wavefunctions, summed over all Fock states (Drell and Yan, 1970, Brodsky and Drell, 1980). In the case of matrix elements of the current $j^+$ in a frame with $q^+ = 0$, only diagonal matrix elements in particle number $n' = n$ are needed. In the nonrelativistic limit one can make contact with the usual formulae for form factors in Schrödinger many-body theory. In the case of inclusive reactions, the hadron and nuclear structure functions are the probability distributions constructed from integrals over the absolute squares $|\psi_n|^2$ summed over $n$. In the far off-shell domain of large parton virtuality, one can use perturbative QCD to derive the asymptotic fall-off of the Fock amplitudes, which then in turn leads to the QCD evolution equations for distribution amplitudes and structure functions. More generally, one can prove factorization theorems for exclusive and inclusive reactions which separate the hard and soft momentum transfer regimes, thus obtaining rigorous predictions for the leading power behavior contributions to large momentum transfer cross sections. One can also compute the far off-shell amplitudes within the light-cone wavefunctions where heavy quark pairs appear in the Fock states. Such states persist over time $\tau \simeq P^+/\mathcal{M}^2$ until they are materialized in the hadron collisions. This leads to a number of novel effects in the hadroproduction of heavy quark hadronic states. See (Brodsky, *et al.*, 1992) for further details. A review of the application of light-cone quantized QCD to exclusive processes is given in (Brodsky and Lepage, 1989).

The light-cone approach to QCD has immediate application to nuclear systems:

1. The formalism provides a covariant many-body description of nuclear systems formally similar to nonrelativistic many-body theory.

2. One can derive rigorous predictions for the leading power-law fall-off of nuclear amplitudes, including the nucleon-nucleon potential, the deuteron form factor, and the distributions of nucleons within nuclei at large momentum fraction. For example, the leading electromagnetic form factor of the deuteron falls as $F_d(Q^2) = f(\alpha_s(Q^2))/(Q^2)^5$, where, asymptotically, $f(\alpha_s(Q^2)) \propto \alpha_s(Q^2)^{5+\gamma}$. The leading anomalous dimension $\gamma$ is computed in (Brodsky, Ji, and Lepage, 1983).

3. In general the six-quark Fock state of the deuteron is a mixture of five different color-singlet states. The dominant color configuration of the six quarks corresponds to the usual proton-neutron bound state. However, as $Q^2$ increases, the deuteron form factor becomes sensitive to deuteron wavefunction configurations where all six quarks overlap within an impact separation $b^{\perp i} < \mathcal{O}(1/Q)$. In the asymptotic domain, all five Fock color-singlet components acquire equal weight; *i.e.*, the deuteron wavefunction becomes 80% "hidden color" at short distances. The derivation of the evolution equation for the deuteron distribution amplitude is given in (Brodsky, Ji, and Lepage, 1983) and (Ji and Brodsky, 1986).

4. QCD predicts that Fock components of a hadron with a small color dipole moment can pass through nuclear matter without interactions (Bertsch, *et al.*, 1981, Brodsky and Mueller, 1988). Thus in the case of large momentum transfer reactions where only small-size valence Fock state configurations enter the hard scattering amplitude, both the initial and final state interactions of the hadron states become negligible. There is now evidence for QCD "color transparency" in exclusive virtual photon $\rho$ production for both nuclear coherent and incoherent reactions in the E665 experiment at Fermilab (Fang, 1993), as well as the original measurement at BNL in quasielastic $pp$ scattering in nuclei (Heppelmann, 1990). The recent NE18 measurement of quasielastic electron-proton scattering at SLAC finds results which do not clearly distinguish between conventional Glauber theory predictions and PQCD color transparency (Makins, 1994).

5. In contrast to color transparency, Fock states with large-scale color configurations strongly interact with high particle number production (Blaettel, *et al.* 1993).

6. The traditional nuclear physics assumption that the nuclear form factor factorizes in the form $F_A(Q^2) = \sum_N F_N(Q^2) F_{N/A}^{\text{body}}(Q^2)$, where $F_N(Q^2)$ is the on-shell nucleon form factor is in general incorrect. The struck nucleon is necessarily off-shell, since it must transmit momentum to align the spectator nucleons along the direction of the recoiling nucleus.

7. Nuclear form factors and scattering amplitudes can be factored in the form given by the reduced amplitude formalism (Brodsky and Chertok, 1976), which follows from the cluster decomposition of the nucleus in the limit of zero nuclear binding. The reduced form factor formalism takes into account the fact that each nucleon in an exclusive nuclear transition typically absorbs momentum $Q_N \simeq Q/N$. Tests of this formalism are discussed in a later section.

8. The use of covariant kinematics leads to a number of striking conclusions for the electromagnetic and weak moments of nucleons and nuclei. For example, magnetic moments cannot be written as the naive sum $\mu = \sum \mu_i$ of the magnetic moments of the constituents, except in the nonrelativistic limit where the radius of the bound state is much larger than its Compton scale: $R_A M_A \gg 1$. The deuteron quadrupole moment is in general nonzero even if the nucleon-nucleon bound state has no D-wave component (Brodsky and Hiller, 1983). Such effects are due to the fact that even "static" moments have to be computed as transitions between states of different momentum $p^\mu$ and $p^\mu + q^\mu$ with $q^\mu \to 0$. Thus one must construct current matrix elements between boosted states. The Wigner boost generates nontrivial corrections to the current interactions of bound systems (Brodsky and Primack, 1969).

9. One can also use light-cone methods to show that the proton's magnetic moment $\mu_p$ and its axial-vector coupling $g_A$ have a relationship independent of the assumed form of the light-cone wavefunction (Brodsky and Schlumpf, 1994). At the physical value of the proton radius computed from the slope of the Dirac form factor, $R_1 = 0.76$ fm, one obtains the experimental values for both $\mu_p$ and $g_A$; the helicity carried by the valence $u$ and $d$ quarks are each reduced by a factor $\simeq 0.75$ relative to their nonrelativistic values. At infinitely small radius $R_p M_p \to 0$, $\mu_p$ becomes equal to the Dirac moment, as demanded by the Drell-Hearn-Gerasimov sum rule (Gerasimov, 1965; Drell and Hearn, 1966). Another surprising fact is that as $R_1 \to 0$, the constituent quark helicities become completely disoriented and $g_A \to 0$. We discuss these features in more detail in the following section.

10. In the case of the deuteron, both the quadrupole and magnetic moments become equal to that of an elementary vector boson in the the Standard Model in the limit $M_d R_d \to 0$. The three form factors of the deuteron have the same ratio as that of the $W$ boson in the Standard Model (Brodsky and Hiller, 1983).

11. The basic amplitude controlling the nuclear force, the nucleon-nucleon scattering amplitude can be systematically analyzed in QCD in terms of basic quark and gluon scattering subprocesses. The high momentum transfer behavior of the amplitude from dimensional counting is $\mathcal{M}_{pp \to pp} \simeq f_{pp \to pp}(t/s)/t^4$ at fixed center of mass angle. A review is given in (Brodsky and Lepage, 1989). The fundamental subprocesses, including pinch contributions (Landshoff, 1974), can be classified as arising from both quark interchange and gluon exchange contributions. In the case of meson-nucleon scattering, the quark exchange graphs (Blankenbecler *et al.*, 1973) can explain virtually all of the observed features of large momentum transfer fixed CM angle scattering distributions and ratios (Carroll, 1992). The connection between Regge behavior and fixed angle scattering in perturbative QCD for quark exchange reactions is discussed in (Brodsky, Tang, and Thorn, 1993). (Sotiropoulos and Sterman, 1994) have shown how one can consistently interpolate from fixed angle scaling behavior to the $1/t^8$ scaling behavior of the elastic cross section in the $s \gg -t$, large $-t$ regime.

12. One of the most striking anomalies in elastic proton-proton scattering is the large spin correlation $A_{NN}$ observed at large angles (Krisch, 1992). At $\sqrt{s} \simeq 5$ GeV, the rate for scattering with incident proton spins parallel and normal to the scattering plane is four times larger than scattering with antiparallel polarization. This phenomena in elastic $pp$ scattering can be explained as the effect due to the onset of charm production in the intermediate state at this energy (Brodsky and de Teramond, 1988). The intermediate state $|uuduudc\bar{c}\rangle$ has odd intrinsic parity and couples to the $J = S = 1$ initial state, thus strongly enhancing scattering when the incident projectile and target protons have their spins parallel and normal to the scattering plane.

13. The simplest form of the nuclear force is the interaction between two heavy quarkonium states, such as the $\Upsilon(b\bar{b})$ and the $J/\psi(c\bar{c})$. Since there are no valence quarks in common, the dominant color-singlet interaction arises simply from the exchange of two or more gluons, the analog of the van der Waals molecular force in QED. In principle, one could measure the interactions of such systems by producing pairs of quarkonia in high energy hadron collisions. The same fundamental QCD van der Waals potential also dominates the interactions of heavy quarkonia with ordinary hadrons and nuclei. As shown in (Luke, Manohar, and Savage, 1992), the small size of the $Q\overline{Q}$ bound state relative to the much larger hadron sizes allows a systematic expansion of the gluonic potential using the operator product potential. The matrix elements of multigluon exchange in the quarkonium state can be computed from nonrelativistic heavy quark theory. The coupling of the scalar part of the interaction to large-size hadrons is rigorously normalized to the mass of the state via the trace anomaly. This attractive potential dominates the interactions at low relative velocity. In this way one establishes that the nuclear force between heavy quarkonia and ordinary nuclei is attractive and sufficiently strong to produce nuclear-bound quarkonium (Brodsky, de Teramond, and Schmidt, 1990).

## MOMENTS OF NUCLEONS AND NUCLEI IN THE LIGHT-CONE FORMALISM

Let us consider an effective three-quark light-cone Fock description of the nucleon in which additional degrees of freedom (including zero modes) are parameterized in an effective potential (Lepage and Brodsky, 1980). After truncation, one could in principle obtain the mass $M$ and light-cone wavefunction of the three-quark bound-states by solving the Hamiltonian eigenvalue problem. It is reasonable to assume that adding more quark and gluonic excitations will only refine this initial approximation (Perry, Harindranath, and Wilson, 1990). In such a theory the constituent quarks will also acquire effective masses and form factors. However, even without explicit solutions, one knows that the helicity and flavor structure of the baryon eigenfunctions will reflect the assumed global SU(6) symmetry and Lorentz invariance of the theory. Since we do not have an explicit representation for the effective potential in the light-cone Hamiltonian $H_{\rm LC}^{\rm effective}$ for three-quarks, we shall proceed by making an ansatz for the momentum space structure of the wavefunction $\Psi$. As we will show below, for a given size of the proton, the predictions and interrelations between observables at $Q^2 = 0$, such as the proton magnetic moment $\mu_p$ and its axial coupling $g_A$, turn out to be essentially independent of the shape of the wavefunction (Brodsky and Schlumpf, 1994).

The light-cone model given in (Schlumpf, 1993) provides a framework for representing the general structure of the effective three-quark wavefunctions for baryons. The wavefunction $\Psi$ is constructed as the product of a momentum wavefunction, which is spherically symmetric and invariant under permutations, and a spin-isospin wave function, which is uniquely determined by SU(6)-symmetry requirements. A Wigner–Melosh (Wigner, 1939; Melosh, 1974) rotation is applied to the spinors, so that the wavefunction of the proton is an eigenfunction of $J$ and $J_z$ in its rest frame (Coester and Polyzou, 1982; Leutwyler and Stern, 1978). To represent the range of uncertainty in the possible form of the momentum wavefunction, we shall choose two simple functions of the invariant mass $\mathcal{M}$ of the quarks:

$$\psi_{\text{H.O.}}(\mathcal{M}^2) = N_{\text{H.O.}} \exp(-\mathcal{M}^2/2\beta^2),$$

$$\psi_{\text{Power}}(\mathcal{M}^2) = N_{\text{Power}}(1 + \mathcal{M}^2/\beta^2)^{-p}$$

where $\beta$ sets the characteristic internal momentum scale. Perturbative QCD predicts a nominal power-law fall off at large $k_\perp$ corresponding to $p = 3.5$ (Lepage and Brodsky, 1980). The Melosh rotation insures that the nucleon has $j = \frac{1}{2}$ in its rest system. It has the matrix representation (Melosh, 1974)

$$R_M(x_i, k_{\perp i}, m) = \frac{m + x_i \mathcal{M} - i\vec{\sigma} \cdot (\vec{n} \times \vec{k}_i)}{\sqrt{(m + x_i \mathcal{M})^2 + \vec{k}_{\perp i}^2}}$$

with $\vec{n} = (0, 0, 1)$, and it becomes the unit matrix if the quarks are collinear $R_M(x_i, 0, m) = 1$. Thus the internal transverse momentum dependence of the light-cone wavefunctions also affects its helicity structure (Brodsky and Primack, 1969).

The Dirac and Pauli form factors $F_1(Q^2)$ and $F_2(Q^2)$ of the nucleons are given by the spin-conserving and the spin-flip vector current $J_V^+$ matrix elements ($Q^2 = -q^2$) (Brodsky and Drell, 1980)

$$F_1(Q^2) = \langle p + q, \uparrow | J_V^+ | p, \uparrow \rangle,$$

$$(Q_1 - iQ_2) F_2(Q^2) = -2M \langle p + q, \uparrow | J_V^+ | p, \downarrow \rangle.$$

We then can calculate the anomalous magnetic moment $a = \lim_{Q^2 \to 0} F_2(Q^2)$. [The total proton magnetic moment is $\mu_p = \frac{e}{2M}(1 + a_p)$.] The same parameters as in (Schlumpf, 1993) are chosen; namely $m = 0.263$ GeV (0.26 GeV) for the up- and down-quark masses, and $\beta = 0.607$ GeV (0.55 GeV) for $\psi_{\text{Power}}$ ($\psi_{\text{H.O.}}$) and $p = 3.5$. The quark currents are taken as elementary currents with Dirac moments $\frac{e_q}{2m_q}$. All of the baryon moments are well-fit if one takes the strange quark mass as 0.38 GeV. With the above values, the proton magnetic moment is 2.81 nuclear magnetons, the neutron magnetic moment is $-1.66$ nuclear magnetons. (The neutron value can be improved by relaxing the assumption of isospin symmetry.) The radius of the proton is 0.76 fm; i.e., $M_p R_1 = 3.63$.

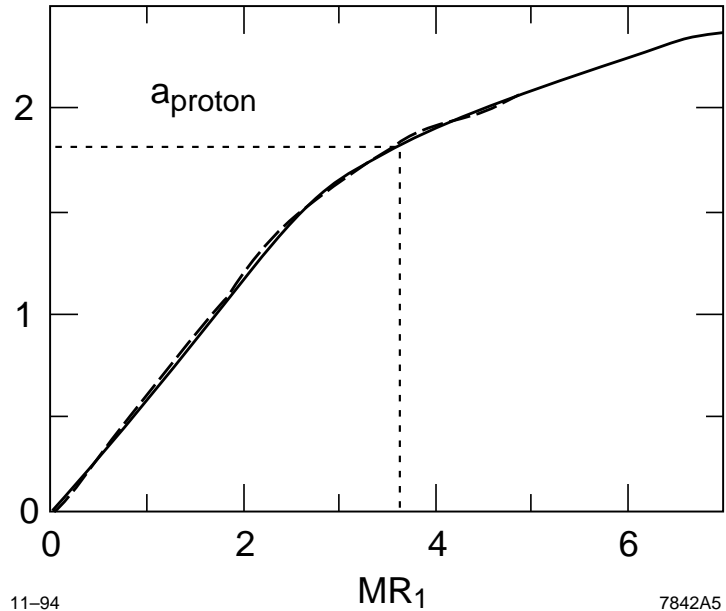

Figure 1. The anomalous magnetic moment $a = F_2(0)$ of the proton as a function of $M_p R_1$: broken line, pole type wavefunction; continuous line, gaussian wavefunction. The experimental value is given by the dotted lines. The prediction of the model is independent of the wavefunction for $Q^2 = 0$.

In Fig. 1 we show the functional relationship between the anomalous moment $a_p$ and its Dirac radius predicted by the three-quark light-cone model. The value of $R_1^2 = -6 dF_1(Q^2)/dQ^2|_{Q^2=0}$ is varied by changing $\beta$ in the light-cone wavefunction while keeping the quark mass $m$ fixed. The prediction for the power-law wavefunction $\psi_{\text{Power}}$ is given by the broken line; the continuous line represents $\psi_{\text{H.O.}}$. Figure 1 shows that when one plots the dimensionless observable $a_p$ against the dimensionless observable $M R_1$ the prediction is essentially independent of the assumed power-law or Gaussian form of the three-quark light-cone wavefunction. Different values of $p > 2$ also do not affect the functional dependence of $a_p(M_p R_1)$ shown in Fig. 1. In this sense the predictions of the three-quark light-cone model relating the $Q^2 \to 0$ observables are essentially model-independent. The only parameter controlling the relation between the dimensionless observables in the light-cone three-quark model is $m/M_p$ which is set to 0.28. For the physical proton radius $M_p R_1 = 3.63$ one obtains the empirical value for $a_p = 1.79$ (indicated by the dotted lines in Fig. 1).

The prediction for the anomalous moment $a$ can be written analytically as $a = \langle \gamma_V \rangle a^{\mathrm{NR}}$, where $a^{\mathrm{NR}} = 2M_p/3m$ is the nonrelativistic ($R \to \infty$) value and $\gamma_V$ is given as (Chung and Coester, 1991)

$$\gamma_V(x_i, k_{\perp i}, m) = \frac{3m}{\mathcal{M}} \left[ \frac{(1-x_3)\mathcal{M}(m + x_3 \mathcal{M}) - \vec{k}_{\perp 3}^2/2}{(m + x_3 \mathcal{M})^2 + \vec{k}_{\perp 3}^2} \right].$$

The expectation value $\langle \gamma_V \rangle$ is evaluated as[*]

$$\langle \gamma_V \rangle = \frac{\int [d^3 k] \gamma_V |\psi|^2}{\int [d^3 k] |\psi|^2}.$$

Let us take a closer look at the two limits $R \to \infty$ and $R \to 0$. In the nonrelativistic limit we let $\beta \to 0$ and keep the quark mass $m$ and the proton mass $M_p$ fixed. In this limit the proton radius $R_1 \to \infty$ and $a_p \to 2M_p/3m = 2.38$ since $\langle \gamma_V \rangle \to 1$[†]. Thus the physical value of the anomalous magnetic moment at the empirical proton radius $M_p R_1 = 3.63$ is reduced by 25% from its nonrelativistic value due to relativistic recoil and nonzero $k_\perp$[‡].

To obtain the ultra-relativistic limit, we let $\beta \to \infty$ while keeping $m$ fixed. In this limit the proton becomes pointlike ($M_p R_1 \to 0$) and the internal transverse momenta $k_\perp \to \infty$. The anomalous magnetic momentum of the proton goes linearly to zero as $a = 0.43 M_p R_1$ since $\langle \gamma_V \rangle \to 0$. Indeed, the Drell-Hearn-Gerasimov sum rule (Gerasimov, 1965; Drell and Hearn, 1966) demands that the proton magnetic moment becomes equal to the Dirac moment at small radius. For a spin-$\frac{1}{2}$ system

$$a^2 = \frac{M^2}{2\pi^2 \alpha} \int_{s_{th}}^{\infty} \frac{ds}{s} \left[ \sigma_P(s) - \sigma_A(s) \right],$$

where $\sigma_{P(A)}$ is the total photoabsorption cross section with parallel (antiparallel) photon and target spins. If we take the point-like limit, such that the threshold for inelastic excitation becomes infinite while the mass of the system is kept finite, the integral over the photoabsorption cross section vanishes and $a = 0$ (Brodsky and Drell, 1980). In contrast, the anomalous magnetic moment of the proton does not vanish in the nonrelativistic quark model as $R \to 0$. The nonrelativistic quark model does not take into account the fact that the magnetic moment of a baryon is derived from lepton scattering at nonzero momentum transfer; *i.e.*, the calculation of a magnetic moment requires knowledge of the boosted wavefunction. The Melosh transformation is also essential for deriving the DHG sum rule and low energy theorems of composite systems (Brodsky and Primack, 1969).

---

[*] $[d^3 k] = d\vec{k}_1 d\vec{k}_2 d\vec{k}_3 \delta(\vec{k}_1 + \vec{k}_2 + \vec{k}_3)$. The third component of $\vec{k}$ is defined as $k_{3i} = \frac{1}{2}(x_i \mathcal{M} - \frac{m^2 + \vec{k}_{\perp i}^2}{x_i \mathcal{M}})$. This measure differs from the usual one used in (Lepage and Brodsky, 1980) by the Jacobian $\prod \frac{dk_{3i}}{dx_i}$ which can be absorbed into the wavefunction.

[†] This differs slightly from the usual nonrelativistic formula $1 + a = \sum_q \frac{e_q}{e} \frac{M_p}{m_q}$ due to the nonvanishing binding energy which results in $M_p \neq 3 m_q$.

[‡] The nonrelativistic value of the neutron magnetic moment is reduced by 31%.

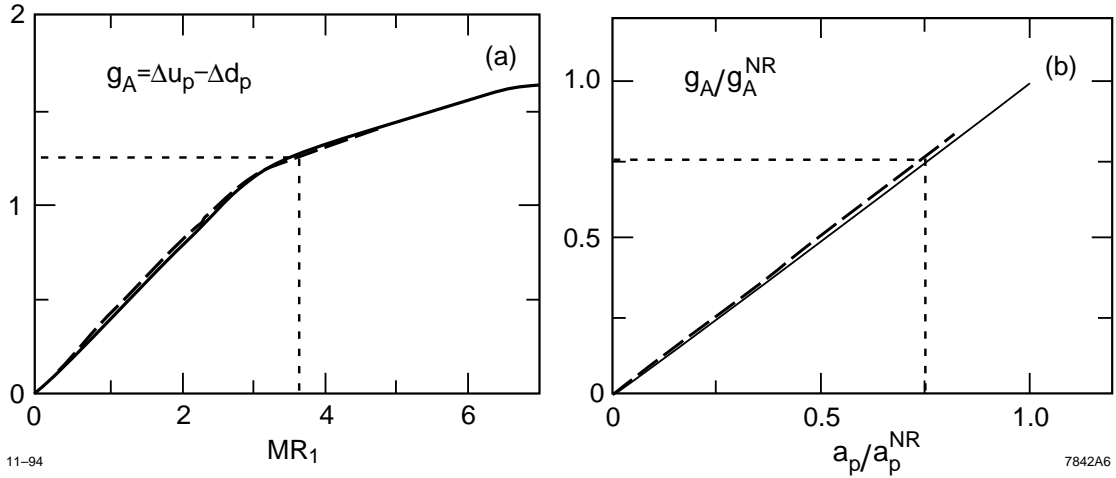

Figure 2. (a) The axial vector coupling $g_A$ of the neutron to proton decay as a function of $M_p R_1$. The experimental value is given by the dotted lines. (b) The ratio $g_A/g_A(R_1 \to \infty)$ versus $a_p/a_p(R_1 \to \infty)$ as a function of the proton radius $R_1$.. The line code is as in Fig. 1.

A similar analysis can be performed for the axial-vector coupling measured in neutron decay. The coupling $g_A$ is given by the spin-conserving axial current $J_A^+$ matrix element $g_A(0) = \langle p, \uparrow | J_A^+ | p, \uparrow \rangle$. The value for $g_A$ can be written as $g_A = \langle \gamma_A \rangle g_A^{\text{NR}}$ with $g_A^{\text{NR}}$ being the nonrelativistic value of $g_A$ and with $\gamma_A$ as (Chung and Coester, 1991; Ma, 1991)

$$\gamma_A(x_i, k_{\perp i}, m) = \frac{(m + x_3 \mathcal{M})^2 - \vec{k}_{\perp 3}^2}{(m + x_3 \mathcal{M})^2 + \vec{k}_{\perp 3}^2}.$$

In Fig. 2(a) the axial-vector coupling is plotted against the proton radius $M_p R_1$. The same parameters and the same line representation as in Fig. 1 are used. The functional dependence of $g_A(M_p R_1)$ is also found to be independent of the assumed wavefunction. At the physical proton radius $M_p R_1 = 3.63$ one predicts the value $g_A = 1.25$ (indicated by the dotted lines in Fig. 2(a)) since $\langle \gamma_A \rangle = 0.75$. The measured value is $g_A = 1.2573 \pm 0.0028$ (Particle Data Group, 1992). This is a 25% reduction compared to the nonrelativistic SU(6) value $g_A = 5/3$, which is only valid for a proton with large radius $R_1 \gg 1/M_p$. As shown in (Ma, 1991), the Melosh rotation generated by the internal transverse momentum spoils the usual identification of the $\gamma^+ \gamma_5$ quark current matrix element with the total rest-frame spin projection $s_z$, thus resulting in a reduction of $g_A$.

Thus, given the empirical values for the proton's anomalous moment $a_p$ and radius $M_p R_1$, its axial-vector coupling is automatically fixed at the value $g_A = 1.25$. This prediction is an essentially model-independent prediction of the three-quark structure of the proton in QCD. The Melosh rotation of the light-cone wavefunction is crucial for reducing the value of the axial coupling from its nonrelativistic value $5/3$ to its empirical value. In Fig. 2(b) we plot $g_A/g_A(R_1 \to \infty)$ versus $a_p/a_p(R_1 \to \infty)$ by varying the proton radius $R_1$. The near equality of these ratios reflects the relativistic spinor structure of the nucleon bound state, which is essentially independent of the detailed shape of the momentum-space dependence of the light-cone wavefunction. We emphasize that at small proton radius the light-cone model predicts not only a vanishing anomalous moment but also $\lim_{R_1 \to 0} g_A(M_p R_1) = 0$. One can understand this physically: in the zero radius limit the internal transverse momenta become infinite and the quark helicities become completely disoriented. This is in contradiction with chiral models which suggest that for a zero radius composite baryon one should obtain the chiral symmetry result $g_A = 1$.

The helicity measures $\Delta u$ and $\Delta d$ of the nucleon each experience the same reduction as $g_A$ due to the Melosh effect. Indeed, the quantity $\Delta q$ is defined by the axial current matrix element

$$\Delta q = \langle p, \uparrow | \bar{q} \gamma^+ \gamma_5 q | p, \uparrow \rangle,$$

and the value for $\Delta q$ can be written analytically as $\Delta q = \langle \gamma_A \rangle \Delta q^{\text{NR}}$ with $\Delta q^{\text{NR}}$ being the nonrelativistic or naive value of $\Delta q$ and with $\gamma_A$.

The light-cone model also predicts that the quark helicity sum $\Delta \Sigma = \Delta u + \Delta d$ vanishes as a function of the proton radius $R_1$. Since the helicity sum $\Delta \Sigma$ depends on the proton size, and thus it cannot be identified as the vector sum of the rest-frame constituent spins. As emphasized in (Ma, 1991), the rest-frame spin sum is not a Lorentz invariant for a composite system. Empirically, one measures $\Delta q$ from the first moment of the leading twist polarized structure function $g_1(x, Q)$. In the light-cone and parton model descriptions, $\Delta q = \int_0^1 dx [q^\uparrow(x) - q^\downarrow(x)]$, where $q^\uparrow(x)$ and $q^\downarrow(x)$ can be interpreted as the probability for finding a quark or antiquark with longitudinal momentum fraction $x$ and polarization parallel or antiparallel to the proton helicity in the proton's infinite momentum frame (Lepage and Brodsky, 1980). [In the infinite momentum there is no distinction between the quark helicity and its spin-projection $s_z$.] Thus $\Delta q$ refers to the difference of helicities at fixed light-cone time or at infinite momentum; it cannot be identified with $q(s_z = +\frac{1}{2}) - q(s_z = -\frac{1}{2})$, the spin carried by each quark flavor in the proton rest frame in the equal time formalism.

Thus the usual SU(6) values $\Delta u^{\text{NR}} = 4/3$ and $\Delta d^{\text{NR}} = -1/3$ are only valid predictions for the proton at large $MR_1$. At the physical radius the quark helicities are reduced by the same ratio 0.75 as $g_A/g_A^{\text{NR}}$ due to the Melosh rotation. Qualitative arguments for such a reduction have been given in (Karl, 1992) and (Fritzsch, 1990). For $M_p R_1 = 3.63$, the three-quark model predicts $\Delta u = 1$, $\Delta d = -1/4$, and $\Delta \Sigma = \Delta u + \Delta d = 0.75$. Although the gluon contribution $\Delta G = 0$ in our model, the general sum rule (Jaffe and Manohar, 1990)

$$\frac{1}{2} \Delta \Sigma + \Delta G + L_z = \frac{1}{2}$$

is still satisfied, since the Melosh transformation effectively contributes to $L_z$.

Suppose one adds polarized gluons to the three-quark light-cone model. Then the flavor-singlet quark-loop radiative corrections to the gluon propagator will give an anomalous contribution $\delta(\Delta q) = -\frac{\alpha_s}{2\pi} \Delta G$ to each light quark helicity (Efremov and Teryaev, 1988). The predicted value of $g_A = \Delta u - \Delta d$ is of course unchanged. For illustration we shall choose $\frac{\alpha_s}{2\pi}\Delta G = 0.15$. The gluon-enhanced quark model then gives the values in Table 1, which agree well with the present experimental values. Note that the gluon anomaly contribution to $\Delta s$ has probably been overestimated here due to the large strange quark mass. One could also envision other sources for this shift of $\Delta q$ such as intrinsic flavor (Fritzsch, 1990). A specific model for the gluon helicity distribution in the nucleon bound state is given in (Brodsky, Burkardt, and Schmidt, 1994).

In summary, we have shown that relativistic effects are crucial for understanding the spin structure of the nucleons. By plotting dimensionless observables against dimensionless observables we obtain model-independent relations independent of the momentum-space form of the three-quark light-cone wavefunctions. For example, the value of $g_A \simeq 1.25$ is correctly predicted from the empirical value of the proton's anomalous moment. For the physical proton radius $M_p R_1 = 3.63$ the inclusion of the Wigner (Melosh) rotation due to the finite relative transverse momenta of the three quarks results in a $\simeq 25\%$ reduction of the nonrelativistic predictions for the anomalous magnetic moment, the axial vector coupling, and the quark helicity content of the proton. At zero radius, the quark helicities become completely disoriented because of the large internal momenta, resulting in the vanishing of $g_A$ and the total quark helicity $\Delta\Sigma$.

Table I

Comparison of the quark content of the proton in the nonrelativistic quark model (NR), in our three-quark model (3q), in a gluon-enhanced three-quark model (3q+g), and with experiment (Ellis and Karliner, 1994).

| Quantity | NR | 3q | 3q+g | Experiment |
|---|---|---|---|---|
| $\Delta u$ | $\frac{4}{3}$ | 1 | 0.85 | $0.83 \pm 0.03$ |
| $\Delta d$ | $-\frac{1}{3}$ | $-\frac{1}{4}$ | $-0.40$ | $-0.43 \pm 0.03$ |
| $\Delta s$ | 0 | 0 | $-0.15$ | $-0.10 \pm 0.03$ |
| $\Delta\Sigma$ | 1 | $\frac{3}{4}$ | 0.30 | $0.31 \pm 0.07$ |

## APPLICATIONS TO NUCLEAR SYSTEMS

We can analyze a nuclear system in the same way as we did the nucleon in the preceding chapter. The triton, for instance, is modeled as a bound state of a proton and two neutrons. The same formulae as in the preceding chapter are valid (for spin-$\frac{1}{2}$ nuclei); we only have to use the appropriate parameters for the constituents.

The light-cone analysis yields nontrivial corrections to the moments of nuclei. For example, consider the anomalous magnetic moment $a_d$ and anomalous quadrupole moment $Q_d^a = Q_d + e/M_d^2$ of the deuteron. As shown in (Tung, 1968), these moments satisfy the sum rule

$$a_d^2 + \frac{2t}{M_d^2}(a_d + \frac{M_d}{2}Q_d^a)^2 = \frac{1}{4\pi}\int_{\nu_{th}^2}^{\infty} \frac{d\nu^2}{(\nu - t/4)^3}(\mathrm{Im} f_P(\nu,t) - \mathrm{Im} f_A(\nu,t)).$$

Here $f_{P(A)}(\nu,t)$ is the non-forward Compton amplitude for incident parallel (antiparallel) photon-deuteron helicities. Thus, in the pointlike limit where the threshold for particle excitation $\nu_{th} \to \infty$, the deuteron acquires the same electromagnetic moments $Q_d^a \to 0, a_d \to 0$ as that of the $W$ in the Standard Model (Brodsky and Hiller, 1983). The approach to zero anomalous magnetic and quadrupole moments for $R_d \to 0$ is shown in Figs. 3 and 4. Thus, even if the deuteron has no D-wave component, a nonzero quadrupole moment arises from the relativistic recoil correction. This correction, which is mandated by relativity, could cure a long-standing discrepancy between experiment and the traditional nuclear physics predictions for the deuteron quadrupole. Conventional nuclear theory predicts a quadrupole moment of 7.233 GeV$^{-2}$ which is smaller than the experimental value $(7.369 \pm 0.039)$ GeV$^{-2}$. The light-cone calculation for a pure S-wave gives a positive contribution of 0.08 GeV$^{-2}$ which accounts for most of the previous discrepancy.

In the case of the tritium nucleus, the value of the Gamow-Teller matrix element can be calculated in the same way as we calculated the axial vector coupling $g_A$ of the nucleon in the previous section. The correction to the nonrelativistic limit for the S-wave contribution is $g_A = \langle \gamma_A \rangle g_A^{\mathrm{NR}}$. For the physical quantities of the triton we get $\langle \gamma_A \rangle = 0.99$. This means that even at the physical radius, we find a nontrivial nonzero correction of order $-0.01$ to $g_A^{\mathrm{triton}}/g_A^{\mathrm{nucleon}}$ due to the relativistic recoil correction implicit in the light-cone formalism. The Gamow-Teller matrix element is measured to be $0.961 \pm 0.003$. The wave function of the tritium ($^3$H)

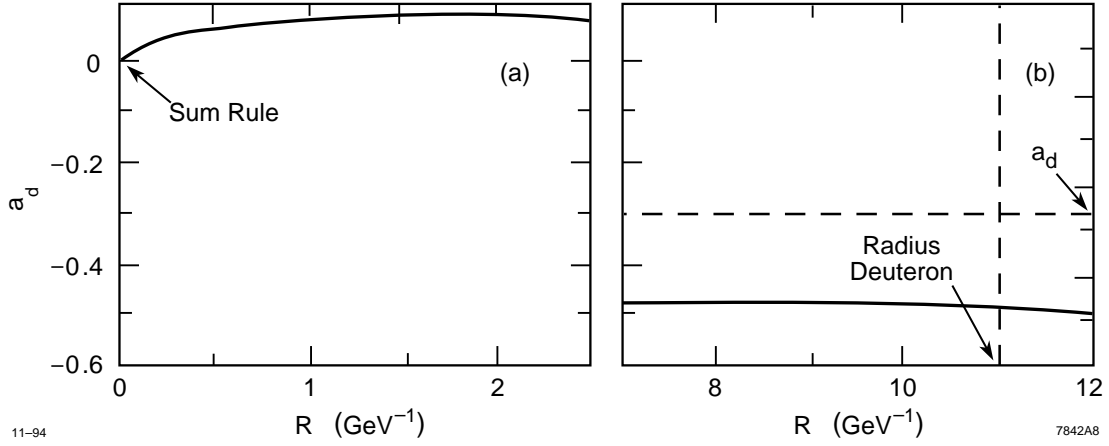

Figure 3. The anomalous moment $a_d$ of the deuteron as a function of the deuteron radius $R_d$. In the limit of zero radius, the anomalous moment vanishes.

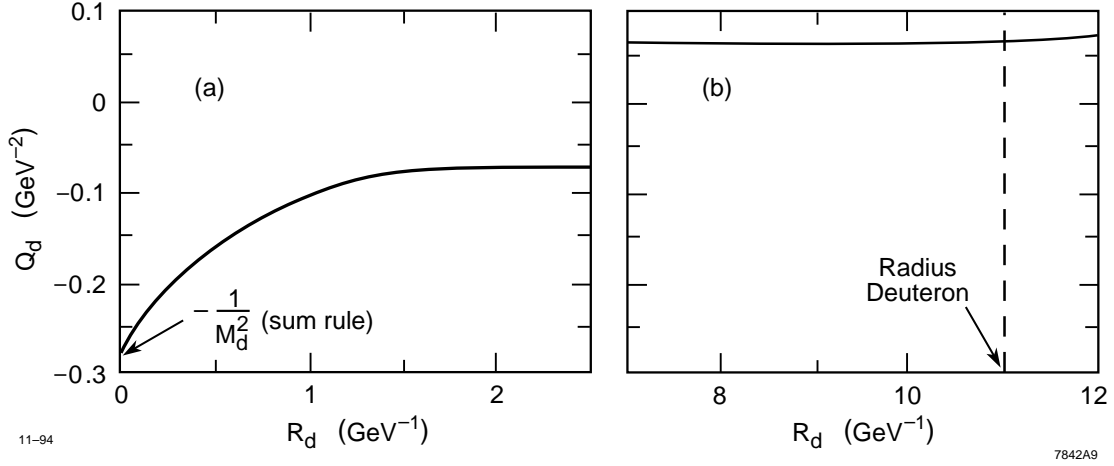

Figure 4. The quadrupole moment $Q_d$ of the deuteron as a function of the deuteron radius $R_d$. In the limit of zero radius, the quadrupole moment approaches its canonical value $Q_d = -e/M_d^2$.

is a superposition of a dominant S-state and small D- and S'-state components $\phi = \phi_S + \phi_{S'} + \phi_D$. The Gamow-Teller matrix element in the nonrelativistic theory is then given by $g_A^{\text{triton}}/g_A^{\text{nucleon}} = (|\phi_S|^2 - \frac{1}{3}|\phi_{S'}|^2 + \frac{1}{3}|\phi_D|^2)(1 + 0.0589) = 0.974$, where the last term is a correction due to meson exchange currents. Figure 5 shows that the Gamow-Teller matrix element of tritium must approach zero in the limit of small nuclear radius, just as in the case of the nucleon as a bound state of three

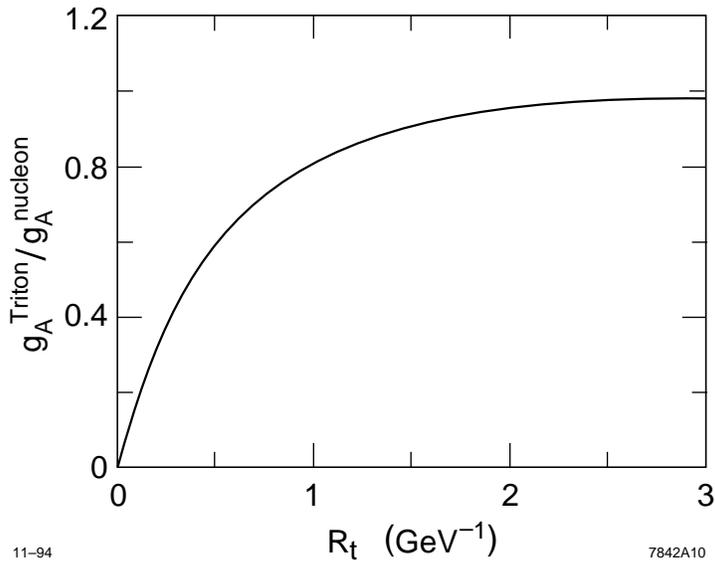

Figure 5. The reduced Gamow-Teller matrix element for tritium decay as a function of the tritium radius.

quarks. This phenomenon is confirmed in the light-cone analysis.

**EXCLUSIVE NUCLEAR PROCESSES**

One of the most elegant areas of application of QCD to nuclear physics is the domain of large momentum transfer exclusive nuclear processes. Rigorous results for the asymptotic properties of the deuteron form factor at large momentum transfer are given in (Brodsky, Ji, and Lepage, 1983). In the asymptotic limit $Q^2 \to \infty$ the deuteron distribution amplitude, which controls large momentum transfer deuteron reactions, becomes fully symmetric among the five possible color-singlet combinations of the six quarks. One can also study the evolution of the "hidden color" components (orthogonal to the $np$ and $\Delta\Delta$ degrees of freedom) from intermediate to large momentum transfer scales; the results also give constraints on the nature of the nuclear force at short distances in QCD. The existence of hidden color degrees of freedom further illustrates the complexity of nuclear systems in QCD. It is conceivable that six-quark $d^*$ resonances corresponding to these new degrees of freedom may be found by careful searches of the $\gamma^* d \to \gamma d$ and $\gamma^* d \to \pi d$ channels.

The basic scaling law for the helicity-conserving deuteron form factor is $F_d(Q^2) \sim 1/Q^{10}$ which comes from simple quark counting rules, as well as perturbative QCD. One cannot expect this asymptotic prediction to become accurate until very large $Q^2$ since the momentum transfer has to be shared by at least six constituents. However, one can identify the QCD physics due to the compositeness of the nucleus, with respect to its nucleon degrees of freedom by using the reduced amplitude formalism (Brodsky and Chertok, 1976). For example, consider the deuteron form factor in QCD. By definition this quantity is the probability amplitude for the deuteron to scatter from $p$ to $p+q$ but remain intact.

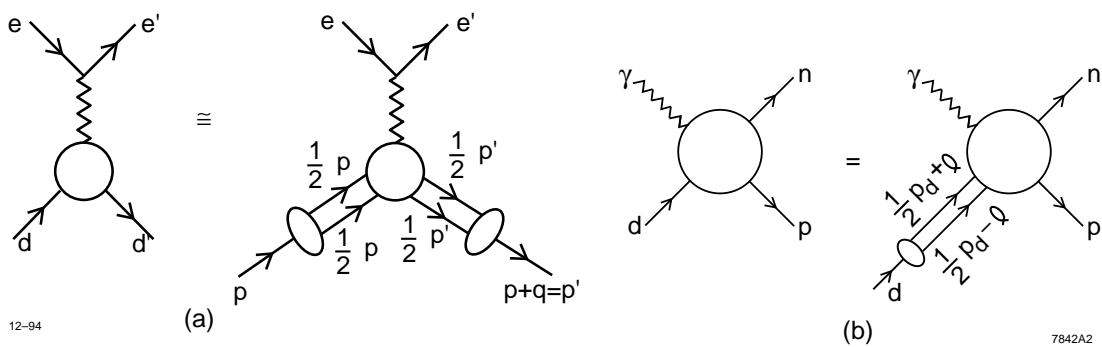

Figure 6. (a) Application of the reduced amplitude formalism to the deuteron form factor at large momentum transfer. (b) Construction of the reduced nuclear amplitude for two-body inelastic deuteron reactions.

Note that for vanishing nuclear binding energy $\epsilon_d \to 0$, the deuteron can be regarded as two nucleons sharing the deuteron four-momentum (see Fig. 6(a)). In the zero-binding limit one can show that the nuclear light-cone wavefunction properly decomposes into a product of uncorrelated nucleon wavefunctions (Ji and Brodsky, 1986). The momentum $\ell$ is limited by the binding and can thus be neglected, and to first approximation, the proton and neutron share the deuteron's momentum equally. Since the deuteron form factor contains the probability amplitudes for the proton and neutron to scatter from $p/2$ to $p/2 + q/2$, it is natural to define the reduced deuteron form factor (Brodsky and Chertok, 1976; Brodsky, Ji, and Lepage, 1983; Ji and Brodsky, 1986):

$$f_d(Q^2) \equiv \frac{F_d(Q^2)}{F_{1N}\left(\frac{Q^2}{4}\right) F_{1N}\left(\frac{Q^2}{4}\right)}.$$

The effect of nucleon compositeness is removed from the reduced form factor. QCD then predicts the scaling

$$f_d(Q^2) \sim \frac{1}{Q^2}$$

*i.e.* the same scaling law as a meson form factor. Diagrammatically, the extra power of $1/Q^2$ comes from the propagator of the struck quark line, the one propagator not contained in the nucleon form factors. Because of hadron helicity conservation, the prediction is for the leading helicity-conserving deuteron form factor ($\lambda = \lambda' = 0$.) As shown in Fig. 7, this scaling is consistent with experiment for $Q = p_T \gtrsim 1$ GeV.

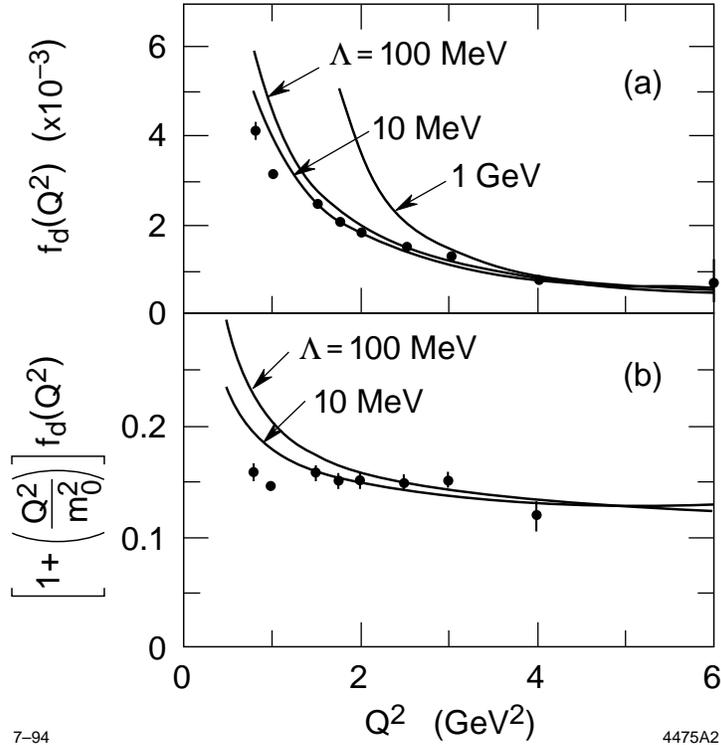

Figure 7. Scaling of the deuteron reduced form factor. The data are summarized in (Brodsky and Hiller, 1983).

The distinction between the QCD and other treatments of nuclear amplitudes is particularly clear in the reaction $\gamma d \rightarrow np$; *i.e.* photo-disintegration of the deuteron at fixed center of mass angle. Using dimensional counting (Brodsky and Farrar, 1975), the leading power-law prediction from QCD is simply $\frac{d\sigma}{dt}(\gamma d \to np) \sim F(\theta_{\rm cm})/s^{11}$. A comparison of the QCD prediction with the recent experiment of (Belz et al., 1994) is shown in Fig. 8, confirming the validity of the QCD scaling prediction up to $E_\gamma \simeq 3$ GeV. One can take into account much of the finite-mass, higher-twist corrections by using the reduced amplitude formalism (Brodsky and Hiller, 1983). The photo-disintegration amplitude contains the probability amplitude (i.e. nucleon form factors) for the proton and neutron to each remain intact after absorbing momentum transfers $p_p - 1/2p_d$ and $p_n - 1/2p_d$, respectively (see Fig. 6(b)). After the form factors are removed, the remaining "reduced" amplitude should scale as $F(\theta_{\rm cm})/p_T$. The single inverse power of transverse momentum $p_T$ is the slowest conceivable in any theory, but it is the unique power predicted by PQCD.

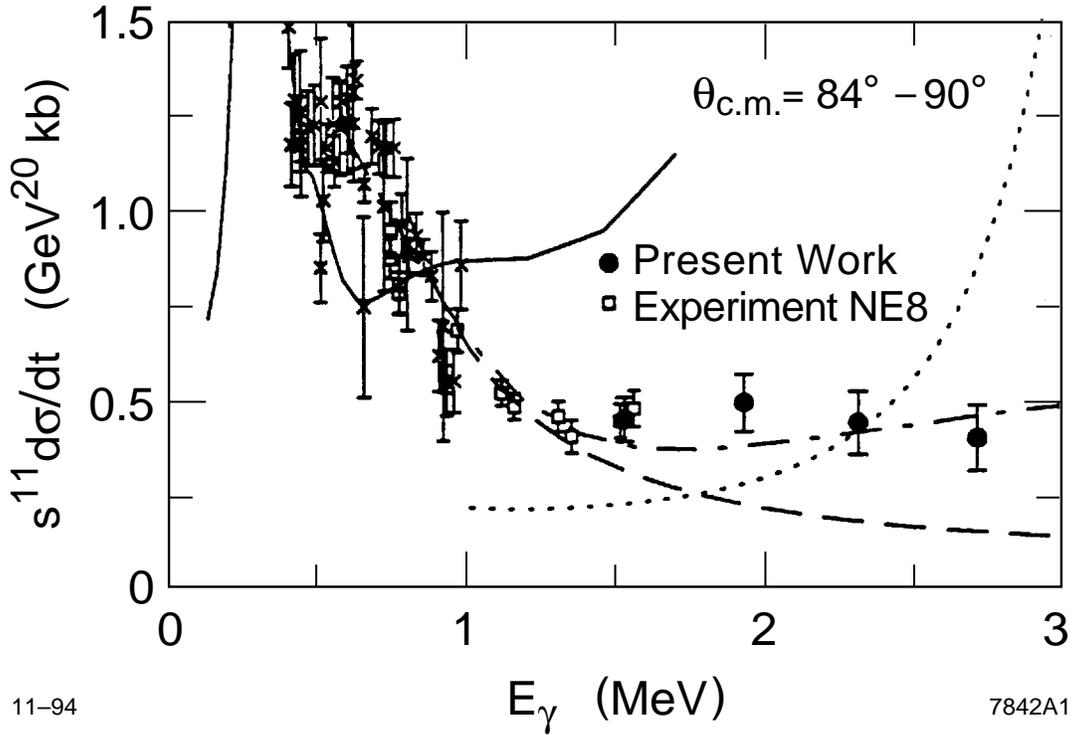

Figure 8. Comparison of deuteron photodisintegration data with the scaling prediction which requires $s^{11}d\sigma/dt(s,\theta_{cm})$ to be at most logarithmically dependent on energy at large momentum transfer. The data and predictions from conventional nuclear theory in are summarized in (Belz et al., 1994).

There are a number of related tests of QCD and reduced amplitudes which require $\overline{p}$ beams (Ji and Brodsky, 1986), such as $\overline{p}d \to \gamma n$ and $\overline{p}d \to \pi p$ in the fixed $\theta_{\rm cm}$ region. These reactions are particularly interesting tests of QCD in nuclei. Dimensional counting rules predict the asymptotic behavior $\frac{d\sigma}{dt}$ ($\overline{p}d \to \pi p$) $\sim \frac{1}{(p_T^2)^{12}} f(\theta_{\rm cm})$ since there are 14 initial and final quanta involved. Again one notes that the $\overline{p}d \to \pi p$ amplitude contains a factor representing the probability amplitude (*i.e.* form factor) for the proton to remain intact after absorbing momentum transfer squared $\hat{t} = (p - 1/2 p_d)^2$ and the $\overline{N}N$ time-like form factor at $\hat{s} = (\overline{p} + 1/2 p_d)^2$. Thus $\mathcal{M}_{\overline{p}d \to \pi p} \sim F_{1N}(\hat{t}) \, F_{1N}(\hat{s}) \, \mathcal{M}_r$, where $\mathcal{M}_r$ has the same QCD scaling properties as quark meson scattering. One thus predicts

$$\frac{\frac{d\sigma}{d\Omega} (\overline{p}d \to \pi p)}{F_{1N}^2(\hat{t}) \, F_{1N}^2(\hat{s})} \sim \frac{f(\Omega)}{p_T^2} \ .$$

**Conclusions**

As we have emphasized in these lectures, QCD and relativistic Fock methods provide a new perspective on nuclear dynamics and properties. In many some cases the covariant approach fundamentally contradicts standard nuclear assumptions. More generally, the synthesis of QCD with the standard nonrelativistic approach can be used to constrain the analytic form and unknown parameters in the conventional theory, as in Bohr's correspondence principle. For example, the reduced amplitude formalism and PQCD scaling laws provide analytic constraints on the nuclear amplitudes and potentials at short distances and large momentum transfers.

Acknowledgments

This work was supported by the Department of Energy, contract DE-AC03-76SF00515. SJB is grateful to Professor Amand Faessler for his kind hospitality at Erice.